\documentclass[12pt,preprint]{aastex}

\shorttitle{New $VR$ magnification ratios of QSO 0957+561}
\shortauthors{Goicoechea et al.}

\begin{document}

\title{New $VR$ magnification ratios of QSO 0957+561}

\author{L.\ J. Goicoechea\altaffilmark{1}, R. Gil--Merino\altaffilmark{1},
A. Ull\'an\altaffilmark{1},M. Serra--Ricart\altaffilmark{2}, J.\ A. 
Mu\~noz\altaffilmark{3}, E. Mediavilla\altaffilmark{2}, 
J. Gonz\'alez--Cadelo\altaffilmark{1}, A. Oscoz\altaffilmark{2}}

\altaffiltext{1}{Departamento de F\'{\i}sica Moderna, Universidad de
Cantabria, Avda. de Los Castros s/n, E-39005 Santander, Cantabria, Spain;
goicol@unican.es, gilmerinor@unican.es, aurora.ullan@postgrado.unican.es,
juan.gonzalezc@alumnos.unican.es}
\altaffiltext{2}{Instituto de Astrof\'{\i}sica de Canarias, C/ V\'{\i}a
L\'actea s/n, E-38205 La Laguna, Tenerife, Spain; mserra@ot.iac.es, 
emg@ll.iac.es, aoscoz@ll.iac.es}
\altaffiltext{3}{Departament d'Astronomia i Astrof\'{\i}sica, Universidad de 
Valencia, Dr. Moliner 50, E-46100 Burjassot, Spain; jmunoz@uv.es}

\begin{abstract}
We present $VR$ magnification ratios of QSO 0957+561, which are inferred from the GLITP light 
curves of Q0957+561A and new frames taken with the 2.56m Nordic Optical Telescope about 14 
months after the GLITP monitoring. To extract the fluxes of the two close components, two 
different photometric techniques are used: {\it pho2comC} and {\it psfphot}. From the two
photometric approaches and a reasonable range for the time delay in the system (415--430 days), 
we do not obtain achromatic optical continuum ratios, but ratios depending on the wavelength. 
Our final global measurements $\Delta m_{AB}(\lambda_V)$ = 0.077 $\pm$ 0.023 mag and $\Delta 
m_{AB}(\lambda_R)$ = 0.022 $\pm$ 0.013 mag (1$\sigma$ intervals) are in agreement with the 
Oslo group results (using the same telescope in the same seasons, but another photometric task
and only one time delay of about 416 days). These new measurements are consistent with 
differential extinction in the lens galaxy, the Lyman limit system, the damped Ly $\alpha$ system,
or the host galaxy of the QSO. The possible values for the differential extinction and the ratio 
of total to selective extinction in the $V$ band are reasonable. Moreover, crude probability 
arguments suggest that the ray paths of the two components cross a similar dusty environment, 
including a network of compact dust clouds and compact dust voids. As an alternative (in fact, the 
usual interpretation of the old ratios), we also try to explain the new ratios as caused by 
gravitational microlensing in the deflector. From magnification maps for each of the gravitationally 
lensed images, using different fractions of the surface mass density represented by the microlenses, 
as well as different sizes and profiles of the $V$--band and $R$--band sources, several synthetic 
distributions of [$\Delta m_{AB}(\lambda_V)$,$\Delta m_{AB}(\lambda_R)$] pairs are derived. In some 
gravitational scenarios, there is an apparent disagreement between the observed pair of ratios and 
the simulated distributions. However, several microlensing pictures work well. To decide between 
either extinction, or microlensing, or a mixed scenario (extinction + microlensing), new 
observational and interpretation efforts are required.
\end{abstract}


\section{Introduction}

The first lensed quasar QSO 0957+561 (Walsh, Carswell, \& Weymann 1979) is an important 
laboratory to study the evolution of multiwavelength magnification ratios
in a lens system and the nature of the involved physical phenomena. However, although a 
lot of magnification ratios were measured during the last twenty years, there is no a 
fair and complete picture accounting for them. The magnification ratio (in magnitudes) is
defined as the difference between the magnitude of Q0957+561A and the time delay 
corrected magnitude of Q0957+561B. At an observed wavelength $\lambda$ and time $t$, the 
ratio is $\Delta m_{AB}(\lambda,t) = m_A(\lambda,t) - m_B(\lambda,t+\Delta t_{AB})$, 
where $\Delta t_{AB}$ is the time delay between the two components. Different studies 
have stablished that the radio magnification ratio does not depend on the time and it is 
close to $-$ 0.31 mag (e.g., Conner, Leh\'ar \& Burke 1992; Garrett et al. 1994). 
Moreover, the optical emission-lines ratio agrees with the radio magnification ratio: 
$\Delta m_{AB}({\rm radio}) \approx \Delta m_{AB}({\rm emission-lines}) \approx -$ 0.31 
mag (e.g., Schild \& Smith 1991). These radio/optical results suggest that the macrolens 
magnification ratio must be of $-$ 0.31 mag. 

On the other hand, the measurements in several optical filters (mostly containing optical 
continuum light) disagree with the macrolens ratio. The $R$--band  magnification ratio 
has been more or less constant for about 15 years, but relatively far from the expected 
ratio for no extinction/microlensing (Pelt et al. 1998; Oscoz et al. 2002; Ovaldsen et al. 
2003a): $0 \leq \Delta m_{AB}(\lambda_R,{\rm 1986-1999}) \leq 0.1$. This ratio was
intensively derived from modest ($\sim$ 1m) telescopes in different observatories. Another 
intriguing result was obtained from the Apache Point Observatory (APO) data. Kundi\'c et al. 
(1997) used the 3.5m telescope at the APO, and estimated magnification ratios of $\Delta 
m_{AB}(\lambda_g,{\rm 1995}) \approx 0.12$ mag and $\Delta m_{AB}(\lambda_r,{\rm 1995}) \approx
0.21$ mag. These last $gr$ ratios were analyzed by Press \& Rybicki (1998), who did not obtain 
any fair conclusion. However we feel that the $r$-band ratio is strongly biased because of the 
methodology to derive it. From the $gr$ light curves, Kundi\'c et al. (1997) simultaneously 
fitted the magnification ratios (magnitude offsets) and the contaminations of the lens galaxy. 
Nevertheless, as the contamination of the lens galaxy in a given optical band is basically a 
noisy magnitud offset, the Kundi\'c et al.'s technique only works when the magnification ratio 
is noticeably larger than the contamination. Thus, whereas the $g$--band ratio may be close to
the true ratio in that optical band, the $r$--band ratio could represent the sum of two 
contributions: a part related to the true magnification ratio and an important contribution due
to the lens galaxy. We note that the intensive $R$--band data agree with this reasonable 
hypothesis. 

Once we {\it solved} the $R-r$ paradox, it seems still necessary to do a detailed 
analysis of the magnification ratios in the red and blue regions of the optical continuum, 
study the possible evolution of the ratios, and try the interpretation of the 
results from different perspectives (e.g., extinction and microlensing). In order to justify 
the optical continuum ratios, gravitational microlensing scenarios are so far the favorite 
ones. Two facts determine this choice. First, radio--signals and optical emission--lines 
reasonably come from regions much larger than the optical continuum sources, so microlensing 
exclusively would change the flux from the optical continuum (compact) sources. Moreover, a 
dusty environment could lead to emission--lines ratios different to $\approx -$ 0.31 mag. 
Second, there is ambiguity on the chromaticity/achromaticity and evolution/rest of the ratios 
(using the oldest data, some authors claimed the existence of a drop in 1983: $\Delta 
m_{AB}(\lambda_R,{\rm 1983}) \sim - 0.3$ mag, which would represent the {\it recuperation} of 
the macrolens magnification ratio and would be difficult to interpret: a void in a dust cloud 
crossing the A image, or an interface between two consecutive dust clouds or microlensing 
events?). This ambiguity permits to assume achromaticity and rest, which are fully consistent 
with sources moving across a roughly constant magnification pattern (on the scale and path of 
the sources). However, Press \& Rybicki (1998) remarked that the {\it excessive constancy} of 
the optical ratios from 1980 through 1995 (and so, the constancy of the magnification) is 
relatively discordant with a microlensing scenario.   
    
In this paper (Section 2) we present new $VR$ magnification ratios of QSO 0957+561. The data 
correspond to the 2000/2001 seasons. In Section 3, the results are compared with the
predictions of a dusty intervening system. In Section 4, we study the feasibility of an 
interpretation based on a population of microlenses within the deflector (lens galaxy + cluster). 
Finally, the main conclusions and future prospects are included in 
Section 5.  

\section{Optical continuum magnification ratios}

We observed QSO 0957+561 from 2000 February 3 to 2000 March 31, as well as in 2001 April. All 
observations were made with the 2.56m Nordic Optical Telescope (NOT) in the $V$ and $R$ bands. 
The main period (in 2000) corresponds to the Gravitational Lenses International Time Project 
(GLITP) monitoring, which was already reduced, analyzed and discussed (Ull\'an et al. 2003). 
In 2001 April, we obtained frames at two nights. Unfortunately, only the frames on April 10 
are useful for photometric tasks. We used the Bessel $V$ and $R$ passbands, with effective 
wavelengths of $\lambda_V$ = 0.536 $\mu$m and $\lambda_R$ = 0.645 $\mu$m, respectively. 
Therefore, as the source quasar is at redshift $z_s$ = 1.41, we deal with the rest-frame 
ultraviolet continuum (0.222--0.268 $\mu$m). 

From the GLITP brightness records of Q0957+561A and the about 14 months delayed Q0957+561B 
fluxes, one can find the $V$--band and $R$--band magnification ratios. Thus, first, we take 
the Q0957+561A fluxes in Figs. (6--7) of Ull\'an et al. (2003). These fluxes were derived from
two photometric approaches: {\it pho2com} and {\it psfphot}, which led to two datasets 
consistent each other. Second, using the {\it pho2com} task, the subtraction of the lens 
galaxy in the Q0957+561B region is not perfect, and some additional correction must be 
incorporated (Serra-Ricart et al. 1999). Ull\'an et al. presented GLITP correction laws that
qualitatively agree with contamination trends for the IAC--80 telescope/camera (operated in the
Teide Observatory by the Instituto de Astrofisica de Canarias), so we use the 
{\it pho2com} method and the GLITP contamination laws to extract clean fluxes of Q0957+561B on 
2001 April 10 (at the time $t_B$). The combined technique ({\it pho2com + corr}) is called {\it 
pho2comC}. Third, we obtain the fluxes of Q0957+561B at $t_B$ through the {\it psfphot} task. 
In this photometric technique, the way to determine the brightnesses of the two quasar 
components and the lens galaxy is from PSF fitting. We use the code developed and described by 
McLeod et al. (1998), which was kindly provided us. Several details on the {\it psfphot} 
approach were included in section 2 of that paper. 
The $V$--band (blue filled circles) and $R$--band (red filled circles) ratios are presented in 
Figure 1. While the top panel contains the results from {\it pho2comC}, the bottom panel includes 
the {\it psfphot} results. The $\Delta m_{AB}$ differences are computed in a simple way: given an 
optical filter, a task and a time delay $\Delta t_{AB}$, we compare the Q0957+561A data in a 
5--days bin centered on $t_B-\Delta t_{AB}$ and the Q0957+561B flux at $t_B$. Only relatively
populated bins are taken into account, i.e., bins with 3 or more data. It is a clear matter that 
the $\Delta m_{AB}(\lambda_V,{\rm 2000}) = \Delta m_{AB}(\lambda_V)$ and $\Delta 
m_{AB}(\lambda_R,{\rm 2000}) = \Delta m_{AB}(\lambda_R)$ are not similar. In other words, given 
a photometric approach and a time delay, the magnification ratios are not achromatic, and it is 
apparent the existence of a correlation between the ratios and the corresponding wavelengths, with 
the higher ratio for the smaller wavelength. 

However, as we only have one frame in both filters ($V$ and $R$) at $t_B$, the $VR$ fluxes of 
Q0957+561B are not so well--determined as the free--of--peaks--of--noise and binned fluxes of 
Q0957+561A. Due to this fact, there is a small discrepancy between the $VR$ ratios from the two 
photometries. Fortunately, Ovaldsen et al. (2003b) reported $VR$ fluxes of QSO 0957+561 through
five nights (one in 2000 January and four in 2001 March) of intensive monitoring at the NOT. 
The Oslo group used the same telescope in the same seasons, although they reduced the frames 
from another method (different to {\it pho2comC} or {\it psfphot}) and only tested the $VR$ 
magnification ratios for a delay of about 416 days. Their typical values $\Delta 
m_{AB}(\lambda_V)$ = 0.065 mag (blue open circles in Fig. 1) and $\Delta m_{AB}(\lambda_R)$ = 
0.026 mag (red open circles in Fig. 1) agree with our results, and roughly represent a gap 
larger than the gap associated with {\it psfphot}, but smaller than the {\it pho2comC} gap. 
Once we confirm the absence of strongly biased results, as the $VR$ estimates for a given 
photometry do not seem to be very sensitive to the delay value, we introduce delay--averaged 
$VR$ ratios. In this first step, the {\it effective} ratios are: $\Delta m_{AB}(\lambda_V)$ = 
0.097 $\pm$ 0.011 mag and $\Delta m_{AB}(\lambda_R)$ = 0.012 $\pm$ 0.010 mag, and $\Delta 
m_{AB}(\lambda_V)$ = 0.057 $\pm$ 0.011 mag and $\Delta m_{AB}(\lambda_R)$ = 0.032 $\pm$ 0.007 
mag, from {\it pho2comC} and {\it psfphot}, respectively. The {\it effective} measurements are 
obtained from the averages of the typical values and errors for the different delays. In a 
second step, taking into account the estimates from the two photometric techniques, we infer 
the final global measurements. These $VR$ measurements are: $\Delta m_{AB}(\lambda_V)$ = 0.077 
$\pm$ 0.023 mag and $\Delta m_{AB}(\lambda_R)$ = 0.022 $\pm$ 0.013 mag (1$\sigma$ intervals). 
Each 1$\sigma$ interval accounts for both the scatter between the typical values and the formal
uncertainty of the photometric methods. The two individual contributions are added in 
quadrature.

\section{Dust system}

We are going to discuss the feasibility of two physical scenarios, which are usually 
introduced to explain optical ratios in disagreement with the expected macrolens ratio.
The relative extinction of Q0957+561A is our first option. This relative extinction may
be caused by dust in an intervening system. What system?. As the Milky Way is a bad candidate 
(the {\it psfphot} measurements and a hypothetical dust system at $z$ = 0 lead to unphysical
extinction parameters), one can discard the differential extinction inside the Galaxy.
However, in QSO 0957+561, we have four more candidates. Lens galaxies can 
generate differential extinction between the components of lensed quasars (e.g., Falco et 
al. 1999, Motta et al. 2002). Thus, the $z$ = 0.36 lens galaxy is an obvious candidate. 
It was discovered by Stockton (1980), and it is a giant elliptical galaxy residing in a 
cluster of galaxies. The light related to the A component and the light associated with the 
B component cross over two galaxy regions that are separated by $\sim$ 20 $h^{-1}$ kpc, 
where the Hubble constant is given by $H$ = 100 $h$ km s$^{-1}$ Mpc$^{-1}$, $\Omega$ = 0.3, 
and $\Lambda$ = 0.7. Besides the lens galaxy, there are two Ly absorption-line systems at 
redshifts $z$ = 1.1249 and $z$ = 1.3911 (e.g., Michalitsianos et al. 1997). The object 
closer to the observer ($z$ = 1.1249) is a Lyman limit system. For this system, spectra of 
QSO 0957+561 showed that the absorption in component A is marginally greater compared with 
component B. Therefore, assuming that more gas implies more dust, a relative extinction of 
Q0957+561A could be produced at $z$ = 1.1249. Michalitsianos et al. (1997) did a detailed 
study of the damped Ly $\alpha$ system at $z$ = 1.3911. In this far object, there is a 
strong evidence for different \ion{H}{1} column densities between the lensed components, 
with the line of sight to image A intersecting a larger column density. The result 
$N_A$(\ion{H}{1}) $>$ $N_B$(\ion{H}{1}) was also claimed by Zuo et al. (1997), who suggested 
the possibility of a differential reddening by dust grains in the damped Ly $\alpha$ absorber. 
At $z$ = 1.3911, the ray paths of the lensed images are separated by $\approx$ 160 $h^{-1}$ 
pc (the separation between the ray paths is estimated from the scheme by Smette et al. 1992,
but using an $\Omega$ = 0.3 and $\Lambda$ = 0.7 cosmology). Apart from the lens galaxy and the
Ly absorption--line systems, one can also consider the source's host galaxy at $z$ = 1.41.

Neglecting gravitational microlensing effects and assuming $R_V$--dependent Cardelli, Clayton, 
\& Mathis (1989) extinction laws which are identical for both components, the dusty scenario 
leads to
\begin{equation}
\Delta m_{AB}(\lambda) = \Delta m_{AB}(\infty) + \Delta E(B - V)[a(x)R_V + b(x)] , 
\end{equation}
where $\Delta m_{AB}(\infty)$ is the macrolens ratio or the ratio at $\lambda = \infty$ 
(in our case, $\Delta m_{AB}(\infty) = -$ 0.31 mag; see Introduction), and $\Delta E(B - 
V)$ is the differential extinction. Here, $R_V = A_V/E(B - V)$ is the ratio of total to 
selective extinction in the $V$ band, and $a(x)$ and $b(x)$ are known functions of $x = 
(1 + z_{dust})/\lambda$ (Cardelli et al. 1989). We take the typical delay--averaged $VR$
magnification ratios from {\it pho2comC} and {\it psfphot}, which are considered as two 
extreme cases: a big gap of 0.085 mag ({\it pho2comC}) and a small gap of 0.025 mag ({\it 
psfphot}). The Ovaldsen et al.'s measurement would correspond to an intermediate case of 
$\Delta m_{AB}(\lambda_V) - \Delta m_{AB}(\lambda_R) \approx$  0.040 mag. For each photometry, 
we estimate the two extinction parameters, $\Delta E(B - V)$ and $R_V$, as a function of the 
redshift of an intervening object, $z_{dust}$. The results are showed in Figure 2. The solid 
line represents the extinction parameters from the {\it pho2comC} ratios, whereas the dashed 
line traces the parameters from the {\it psfphot} ratios. In Fig. 2, the red points are the 
solutions for $z_{dust}$ = 0.36 (lens galaxy), the green points are the solutions for 
$z_{dust}$ = 1.1249 (Lyman limit system), the blue points are associated with $z_{dust}$ = 
1.3911 (damped Ly $\alpha$ system), and the yellow points are associated with $z_{dust}$ = 
1.41 (host galaxy). If we focus on the four candidates (lens galaxy, Ly absorption--line 
systems, and QSO's host galaxy), global intervals 
$\Delta E(B - V)$ = 0.03--0.10 mag and $R_V$ = 2--9 are derived. In the Galaxy, typical paths 
have a ratio of total to selective extinction $R_V \approx$ 3.1, while paths through typical 
and non--typical extinction regions have values in the range 2--6. The $R_V$ values in far 
galaxies can be even higher. For example, the $z$ = 0.68 spiral lens B0218+357 has $R_V 
\approx$ 7--12 (Falco et al. 1999; Mu{\~n}oz et al. 2004). Therefore, the results on $R_V$ 
seem reasonable. On the other hand, the range for differential extinction agrees with typical 
values in lens systems, and as a general conclusion, the observations can be explained by 
means of a dusty scenario.

How is the structure of the dust system?. The ray paths of the two components cross 
either different environments (i.e., B crosses an empty zone and A crosses a dusty 
zone) or the same dusty environment. In the first case, as the emission--lines sources do 
not experience (important) differential extinction and the optical continuum compact sources 
suffer it, the A zone would include a compact dust cloud in the trajectory of the A beam 
together with a random distribution of similar clumps of dust. The distribution of clouds cannot
significantly perturb the radiation from the emission--lines sources. If the projected 
(into the source plane) cloud radius is $R_{cloud}$ = 0.01 pc, i.e., similar to the expected 
radial size of the whole accretion disk around a 10$^8$ M$_\odot$ black hole, then the time of 
cloud crossing by the optical continuum region (OCR) is of about 30 years (considering $R_{OCR} 
< R_{cloud}$ and an effective transverse velocity of 600 km s$^{-1}$; e.g., Wambsganss et al. 
2000). If we also consider the existence of $N$ dust clouds in front of the broad--line region 
(BLR) as well as a typical radius $R_{BLR}$ = 0.1 pc, the surface covering factor of the dust 
would be $f_{dust/BLR} = N(R_{cloud}/R_{BLR})^2 = 10^{-2} N$. As $f_{dust/BLR}$ should be much 
smaller than 1 ($f_{dust/BLR} <<$ 1), a constraint on the surface density of clouds can be derived 
in a simple way. It must be much smaller than 10$^4$/$\pi$ clouds pc$^{-2}$, so the probability to 
see a cloud crossing phenomenon is equal to or less than 10\%. The situation is better in the 
second case, when the A and B zones are within the same environment. While a homogeneous 
distribution of dust is not consistent with the optical continuum ratios, an inhomogeneous 
distribution works well. We assume an environment constituted by a network of compact dust clouds 
and compact empty regions ($R_{cloud}$ = $R_{void}$ = 0.01 pc), with a cloud in the trajectory of 
the A beam and a void in the trajectory of the other (B) beam. In this picture, the optical continuum 
ratios would be anomalous, but the emission--line ratios would be very close to the macrolens ratio, 
because two large regions contain a very similar amount of dust. Moreover, both the probability of a 
cloud crossing (component A) and the probability of a void crossing (component B) are of about 50\%. 
Hence, the probability of a cloud(A)/void(B) crossing is crudely of 25\%. For the farest candidate 
(host galaxy), the common environment is a very reasonable approach, and for the nearest candidate 
(lens galaxy), the common environment could be due to the presence of an inhomogeneous dust lane with 
a length of $\geq$ 20 kpc. 

\section{Microlensing}

To explain the chromatic ratios, the gravitational microlensing of QSO 0957+561 is our
second option. While the emission-lines regions are not magnified from gravitational 
microlensing, the optical continuum compact sources could be affected by this physical
phenomenon. Using the values for the normalized surface mass density (convergence) and 
the shear, we produce 2--dimensional magnification maps with a ray shooting technique 
(Wambsganss 1990, 1999). These values are: $\kappa_A$ = 0.22, $\gamma_A$ = 0.17, $\kappa_B$ 
= 1.24, and $\gamma_B$ = 0.9 (Pelt et al. 1998). With respect to the fraction of the surface 
mass density represented by the microlenses, we consider three cases: (a) all the mass is 
assumed to be in granular form (compact objects), (b) 50\% of the mass in granular form, and 
(c) the mass is dominated by a smoothly distributed material, with only 25\% of the mass in 
compact objects. All the microlensing objects are assumed to have a similar mass $M$ and are 
distributed randomly over the lens plane. We obtain detailed results for $M = 1 M_\odot$, and 
comment the expected results for smaller masses ($M < 1 M_\odot$). The effect of the 
source is taken into account by convolving a selected source intensity profile with the 
magnification maps. We consider two different circularly--symmetric profiles: a $p$ = 3/2 
power--law (PL) profile and a Gaussian (GS) profile. The PL profile was introduced by Shalyapin 
(2001), and it is roughly consistent with the standard accretion disk. The GS profile is not so 
smooth as the PL one, and it was used in almost all previous microlensing studies in QSO 
0957+561. In our model we assume that the $V$--band source has a characteristic length 
(intensity distribution) of $R_V$ = 
3 $\times$ 10$^{14}$ cm (small source), $R_V$ = 10$^{15}$ cm (intermediate source), or $R_V$ = 
3 $\times$ 10$^{15}$ cm (large source), and the source size ratio $q = R_V/R_R$ is given by 
about 0.8 or 1/3. For the standard accretion disk model, one has the relationship $R(\lambda) 
\propto \lambda^{4/3}$ (e.g., Shalyapin et al. 2002), and consequently, the standard ratio $q 
\approx$ 0.8, which is close to the critical ratio ($q$ = 1). It is also assumed that the source, 
the deflector and the observer are embedded in a standard flat cosmology: $H$ = 100 $h$ km 
s$^{-1}$ Mpc$^{-1}$, $h$ = 2/3, $\Omega$ = 0.3, $\Lambda$ = 0.7.

In other recent analyses about QSO 0957+561 microlensing (Refsdal et al. 2000; Wambsganss et 
al. 2000), the authors concentrated on the time evolution of $\Delta m_{AB}$ in a given 
optical band. Refsdal et al. (2000) used a GS source and found 90\% confidence level 
constraints for the microlenses mass $M$ and the source radius $R$: $M$ = 2 $\times$ 10$^{-3}$
-- 0.5 $M_\odot$, $R <$ 6 $\times$ 10$^{15}$ cm. On the other hand, Wambsganss et al. (2000) 
claimed that the microlenses mass must be $\geq$ 0.1 $M_\odot$ for small GS sources (3$\sigma$ 
limit). All these previous constraints are based on the hypothesis of a Gaussian source, and 
moreover, as the analyses were focused on the time domain, some hypotheses about both the 
effective transverse velocity and the random microlens motion were needed. However, our 
approach is different, because we wish to study the spectral evolution of $\Delta m_{AB}$ at a 
given emission time. In this paper, we discuss both the GS and PL profiles and do not need to 
introduce a dynamical hypothesis. In fact, in relation to multiepoch simulations, multiband 
simulations have the advantage of being free from assumptions on the peculiar motions of the
quasar and deflector, and the orbits of the microlenses in the lens galaxy and cluster centre. 

The key idea is that the $VR$ magnification ratios are generated by two circular concentric 
sources with different size. The common center of the $V$--band and $R$--band sources is 
placed at two arbitrary pixels of the two magnification patterns, and we look for the 
corresponding [$\Delta m_{AB}(\lambda_V)$,$\Delta m_{AB}(\lambda_R)$] pair. For a given set
of physical parameters, we can test a very large number of pairs of pixels (10$^4$), and thus, 
obtain a distribution of pairs of magnification ratios. In the [$\Delta m_{AB}(\lambda_V)$,$\Delta 
m_{AB}(\lambda_R)$] plane, for $q$ = 1 (critical source size ratio), the simulated distributions 
would lie exactly on the $\Delta m_{AB}(\lambda_R)$ = $\Delta m_{AB}(\lambda_V)$ straight line
(critical line). However, when $q \neq$ 1, we must infer 2D distributions, including 
points at non-zero distances (in magnitudes) from the critical line. The 2D distributions in the 
six panels of Fig. 3 have been derived by using a deflector with only granular matter ($M = 1 
M_\odot$) and GS sources. In Fig. 3 (and also Fig.4), the results for the small, intermediate and 
large $V$--band sources are depicted in the top, middle and bottom panels, respectively. The left 
panels correspond to a standard $q$, and the right panels correspond to a non--standard $q$ ($q$ 
= 1/3). We also draw the rectangle associated with the 1$\sigma$ measurements. It is apparent that 
larger differences between $R_V$ and $R_R$ imply wider distributions. In Fig. 3, we only see a 
very interesting region of the whole plane, which includes the observed magnification ratios. The 
reader can extrapolate the six behaviours in this region towards a larger region along and around 
the critical line. In order to get a good resolution, instead of the whole 2D distributions, we 
decided in favour of a smaller (but representative) region. What about the PL intensity profile?. 
In Fig. 4, we again consider all the mass in compact objects of $M = 1 M_\odot$, but PL sources. 
The new distributions are different to their partners in Fig. 3 (with similar values of $R_V$ and 
$R_R$). However, for obtaining quantitative conclusions, some additional analyses are needful.

We make 12 histograms $N(d)$ for the 12 scenarios where 100\% of the mass is contained in 
compact objects ($M = 1 M_\odot$). Each histogram represents the number of points at different 
distances $d$ (in magnitudes) from the critical line. Given the [$\Delta m_{AB}(\lambda_V)$,$\Delta 
m_{AB}(\lambda_R)$] pair, the distance from the critical line, $d = [\Delta m_{AB}(\lambda_V) - 
\Delta m_{AB}(\lambda_R)]/\sqrt{2}$, and the difference between ratios, $D = \Delta m_{AB}(\lambda_V) 
- \Delta m_{AB}(\lambda_R)$, differ only in a factor $\sqrt{2}$.
Therefore, we analyze histograms $N(d)$, which are basically $D$ distributions. The points above the 
critical line have negative distances ($d <$ 0), whereas the points below the critical line have 
positive distances ($d >$ 0). All the histograms are showed in Fig. 5: GS profile and standard $q$ 
(top left panel), GS profile and $q$ = 1/3 (top right panel), PL profile and standard $q$ (bottom 
left panel), and PL profile and $q$ = 1/3 (bottom right panel). We use blue lines for the small 
$V$--band source, green lines for the intermediate $V$--band source, and red lines for the large 
$V$--band source. The four arrows indicate the distances from the critical line to the four vertexes 
of the 1$\sigma$ rectangle in Figs. 3 and 4. If the true [$\Delta m_{AB}(\lambda_V)$,$\Delta 
m_{AB}(\lambda_R)$] pair would be within the 1$\sigma$ box, then the true distance would verify $d 
\geq$ 0.013 mag. Here, we assume that the true pair is placed at $d \geq$ 0.013 mag, so probabilities 
$P$($d <$ 0.013 mag) are computed. Some probability distributions (histograms) lead to very high values 
of $P$($d <$ 0.013 mag), and thus, we can rule out the corresponding scenarios. On the contrary, other 
$N-d$ relationships are relatively consistent with the constraint. The estimates of $P$($d <$ 0.013 mag)
appear in Table 1, and we see several values $\leq$ 90\%, which suggest {\it realistic} physical 
parameters, or in other words, viable scenarios. In the GS profile case, the feasible pictures 
are related to either the large $V$--band source or an intermediate $V$--band source together
with a non--standard $R$--band companion. In the PL profile case, from the $P \leq$ 90\% criterion, 
all the scenarios seem to be plausible. 

In Fig. 6, we see the magnification maps for the Q0957+561A (left panel) and Q0957+561B (right panel) 
components as well as the 10000 pairs of pixels (white marks) used to produce Figs. 3, 4 and 5. For the 
B component, the granular mass density is high and the caustic network almost takes up the whole map. The 
10$^4$ highlighted pixels in each map are randomly distributed, so the sampling is not biased. In Fig. 7, 
the chromatic microlensing is apparent. The caustic maps in Fig. 6 are convolved with the largest Gaussian 
sources ($R_V$ = 3 $\times$ 10$^{15}$ cm, $R_R$ = 3 $\times$ $R_V$). In the three top panels of Fig. 7, 
we include the $V$--convolved map for the A component (left), the $R$--convolved map for the A component
(centre), and the difference between the two convolutions (right). The results for the B component appear
in the three bottom panels of Fig. 7: $V$--convolution (left), $R$--convolution (centre), and difference
(right). It is clear that the convolutions with the $V$--band source are different to the convolutions
with the $R$--band source. This fact gives rise to the appearance of chromatic effects, i.e., there are
difference signals in the two right panels.

When all the mass is not contained in compact objects, i.e., a certain mass fraction is smoothly 
distributed, we can repeat the above--mentioned procedure and obtain new 2D distributions, histograms,
probabilities, etc. To avoid an "excess of complementary material" (figures), only the $P$($d <$ 
0.013 mag) values are included in Table 1. Two different cases are analyzed in this paper: 50\% of 
the mass in granular form (50\% of the mass in a smoothly distributed component), and 25\% of the
mass in compact objects (75\% of the mass is contained in a smoothly distributed material). As the
percentage of mass in compact objects decreases, it is more difficult to find scenarios in agreement
with the observational constraint. In fact, when the mass is dominated by a smoothly distributed 
material, almost all the physical pictures are ruled out by the observations at the 90\% level. 
Only two rare scenarios (where the source sizes are in a relation 1:3) lead to $P$($d <$ 0.013 mag) 
$\leq$ 90\%, or equivalently, $P$($d \geq$ 0.013 mag) $>$ 10\%. Although from a population of 
microlenses of $M = 1 M_\odot$ we are able to test an important set of physical situations, we
must also think about the expected results from a population of microlenses of $M < 1 M_\odot$. In the
extreme (but still plausible) case of $M = 10^{-2} M_\odot$, the physical size of the magnification
maps (the maps are 2048 pixels a side which cover a physical size of 16 Einstein radii) reduces in 
a factor 10 (just the factor that relates the Einstein radii for 1 $M_\odot$ and $10^{-2} M_\odot$).
Now, the $V$--band and $R$--band sources with radii $R_R$ and $R_V$ work as the sources with radii 
10 $\times R_R$ and 10 $\times R_V$ in the old case ($M = 1 M_\odot$). For example, the results 
in Table 1 for $R_V$ = 3 $\times$ 10$^{15}$ cm would be the results for $R_V$ = 3 $\times$ 10$^{14}$ 
cm and $M = 10^{-2} M_\odot$. 

We remark that the discussion is based on arrays of 2048 by 2048 pixels, which cover 16 $R_E$ 
$\times$ 16 $R_E$ regions. Similar arrays were used by Schmidt \& Wambsganss (1998) in the 
cases $M \geq 10^{-2} M_\odot$, and we think that our results are reliable. However, in 
very detailed studies to discriminate between different microlensing pictures, magnification 
maps with larger spatial coverage and better spatial resolution may be useful. Some of these 
improvements are related to the availability of high performance computers.
   
\section{Conclusions and future work}

We present in this work a very robust estimation of the $VR$ magnification ratios of QSO 0957+561. 
The new measurements are based on observations (in 2000--2001) with the 2.56m Nordic Optical Telescope, 
two different photometric techniques, and a reasonable interval for the time delay in the system. Our 
1$\sigma$ estimates $\Delta m_{AB}(\lambda_V)$ = 0.077 $\pm$ 0.023 mag and $\Delta m_{AB}(\lambda_R)$ = 
0.022 $\pm$ 0.013 mag are supported by the results in Ovaldsen et al. (2003b). A first conclusion is the 
relation $\Delta m_{AB}(\lambda_V) > \Delta m_{AB}(\lambda_R)$, which means that the optical continuum 
magnification ratios are not achromatic, and moreover, the higher ratio corresponds to the smaller 
wavelength. 

To explain the measurements, two alternatives are explored: a dust system between the quasar and the 
observer, or a population of microlenses in the deflector (the most popular picture). We find that 
the observed ratios are consistent with both alternatives, i.e., differential extinction in a dust
system and gravitational microlensing in the deflector. We cannot rule out the existence of a dusty 
scenario constituted by a network of compact dust clouds and compact empty regions, with a cloud in 
the line of sight to the Q0957+561A component and a void in the trajectory of the Q0957+561B beam.
The new estimates cover a narrow spectral range (0.536--0.645 $\mu$m), and this fact plays a dramatic 
role in the discrimination of possible physical scenarios. We think that future accurate measurements 
at a large collection of wavelengths, with a good coverage of the 0.2--1 $\mu$m range, will decide on 
the feasibility of a dust system in QSO 0957+561 and other things. For example, the GTC (a 
segmented 10.4 meter telescope to be installed in one of the best sites of
the Northern Hemisphere: the Roque de los Muchachos Observatory, La Palma, 
Canary Islands, Spain. First light is planed for 2005) 
will have an instrument (OSIRIS) with very interesting preformances. OSIRIS 
will incorporate tunable filters with FWHM from 10$^{-4}$ to 7 $\times$ 10$^{-3}$ $\mu$m over the whole 
optical wavelength range, and thus, tunable imaging of the two quasar components at two epochs (separated 
by the time delay) could lead to solve the current puzzle (dust or microlenses?) and infer tight 
constraints on the physical parameters of the favoured alternative. Observations in the 
ultraviolet wavelength range ($\leq 0.36 \mu$m) may also be a decisive tool. In particular, 
if the actual origin of the magnification ratio anomalies is a dust system in the lens galaxy.

If the 0.2175 $\mu$m extinction bump would be unambiguously detected (assuming dusty scenarios
in the 0.36--1.41 redshift interval, the rest-frame wavelength of the peak corresponds to 
0.30--0.52 $\mu$m in the observer frame), then the current extinction/microlensing degeneration 
would be broken. In this last case (detection of extinction), one may try to derive a very rich information: 
the differential extinction, the ratio of total to selective extinction in the $V$ band, the macrolens ratio, 
the redshift of the dust, and so on (e.g., Falco et al. 1999). On the other hand, if the dusty scenarios 
would be ruled out from future observations at lots of wavelengths, the multiwavelength ratios may be used 
to decide on the viable gravitational microlensing scenarios. From a simple analysis, we show that several 
microlensing scenarios are ruled out by the new observations at the 90\% level (see Table 1 and Sect. 4), so 
future detailed data could lead to important constraints. The future observational work will require new 
ingredients in the microlensing scheme, e.g., new source intensity profiles (in particular, the standard 
accretion disk profile; Shalyapin et al. 2002), or a two mass population of microlenses, describing the 
granular matter in the lens galaxy and the granular matter in the cluster centre (e.g., Schechter, Wambsganss 
\& Lewis 2004). Sophisticated source models could be also included. The double ring structure suggested by
Schild \& Vakulik (2003) is an obvious candidate. These authors have described the $R$--band source, but they
did not quote the chromatic behaviour of the five model parameters: the radius of the large ring, the radial
thickness of the large ring, the radius of the small ring, the thickness of the small ring, and the brightness
ratio of the two rings, which is a basic issue to simulate multiwavelength magnification ratios. 
The extinction and microlensing hypotheses can be tested in both the frequency and time 
domains, and several previous efforts focused on the possible microlensing variability (e.g., Refsdal et al. 
2000; Wambsganss et al. 2000). Obviously, the observations in the time domain (evolution of the magnification 
ratios) have a great interest. However, when they are compared to microlensing simulations, the microlensing 
framework must incorporate dynamical information, i.e., the confrontation between the observed evolution and 
the microlensing scheme requires additional free parameters (which are needless in multiwavelength studies). 
Finally, we note that the two alternative schemes (only dust or only microlenses) might be biased models of the 
actual situation, and a mixed scenario (extinction + microlensing) is also possible. Very recently, to explain 
spectrophotometric observations of the lens system HE 0512$-$3329, Wucknitz et al. (2003) proposed this mixed 
model. 

\acknowledgments

We thank Vyacheslav Shalyapin and an anonymous referee for interesting comments on a first version of the 
paper. The authors would like to acknowledge Joachim Wambsganss for providing his microlensing code. We also 
thank Jorge Casares and the NOT staff for the frames taken at the beginning of April 2001 and 2001 April 10, 
respectively. The observations were made with the Nordic Optical Telescope (NOT), operated on the island of 
La Palma, jointly by Denmark, Finland, Iceland, Norway, and Sweden, in the Spanish Observatorio del Roque de 
los Muchachos of the Instituto de Astrofisica de Canarias. This work was supported by the P6/88 project of 
the IAC, Universidad de Cantabria funds, and the MCyT grants AYA2000-2111-E and AYA2001-1647-C02.

\clearpage

\begin{table}
\begin{center}
\caption{Probabilities for measuring a pair of magnification ratios (one in the $V$ band 
and the other in the $R$ band) at a distance $<$ 0.013 magnitudes from the critical line 
$\Delta m_{AB}(\lambda_R)$ = $\Delta m_{AB}(\lambda_V)$.\label{tbl-1}}
\scriptsize
\begin{tabular}{cccccc}
\tableline\tableline
Mass in granular form\tablenotemark{a} (\%) & Intensity profile & $R_V$ (cm) & $q = R_V/R_R$ 
& $P$($d <$ 0.013 mag) (\%) & $P \leq$ 90\%? \\
\tableline
100 & Gaussian & 3 $\times$ 10$^{14}$ & $\approx$ 0.8 & 93.5 & -- \\
    &          &                      & 1/3 & 91.4 & -- \\
    &          & 10$^{15}$ & $\approx$ 0.8 & 91.1 & -- \\
    &          &           & 1/3 & 81.7 & yes \\
    &          & 3 $\times$ 10$^{15}$ & $\approx$ 0.8 & 80.0 & yes \\
    &          &                      & 1/3 & 68.6 & yes \\
    & Power--law ($p$ = 3/2) & 3 $\times$ 10$^{14}$ & $\approx$ 0.8 & 89.1 & yes \\
    &                        &                      & 1/3 & 85.5 & yes \\
    &                        & 10$^{15}$ & $\approx$ 0.8 & 86.6 & yes \\
    &                        &           & 1/3 & 75.3 & yes \\
    &                        & 3 $\times$ 10$^{15}$ & $\approx$ 0.8 & 78.2 & yes \\
    &                        &                      & 1/3 & 69.0 & yes \\
50  & Gaussian & 3 $\times$ 10$^{14}$ & $\approx$ 0.8 & 97.9 & -- \\
    &          &                      & 1/3 & 96.5 & -- \\
    &          & 10$^{15}$ & $\approx$ 0.8 & 96.8 & -- \\
    &          &           & 1/3 & 90.8 & -- \\
    &          & 3 $\times$ 10$^{15}$ & $\approx$ 0.8 & 91.3 & -- \\
    &          &                      & 1/3 & 80.4 & yes \\
    & Power--law ($p$ = 3/2) & 3 $\times$ 10$^{14}$ & $\approx$ 0.8 & 95.6 & -- \\
    &                        &                      & 1/3 & 93.1 & -- \\
    &                        & 10$^{15}$ & $\approx$ 0.8 & 93.9 & -- \\
    &                        &           & 1/3 & 84.7 & yes \\
    &                        & 3 $\times$ 10$^{15}$ & $\approx$ 0.8 & 85.9 & yes \\
    &                        &                      & 1/3 & 75.8 & yes \\
25  & Gaussian & 3 $\times$ 10$^{14}$ & $\approx$ 0.8 & 99.3 & -- \\
    &          &                      & 1/3 & 99.0 & -- \\
    &          & 10$^{15}$ & $\approx$ 0.8 & 98.8 & -- \\
    &          &           & 1/3 & 96.6 & -- \\
    &          & 3 $\times$ 10$^{15}$ & $\approx$ 0.8 & 95.8 & -- \\
    &          &                      & 1/3 & 89.4 & yes \\
    & Power--law ($p$ = 3/2) & 3 $\times$ 10$^{14}$ & $\approx$ 0.8 & 98.2 & -- \\
    &                        &                      & 1/3 & 97.2 & -- \\
    &                        & 10$^{15}$ & $\approx$ 0.8 & 97.7 & -- \\
    &                        &           & 1/3 & 92.5 & -- \\
    &                        & 3 $\times$ 10$^{15}$ & $\approx$ 0.8 & 92.3 & -- \\
    &                        &                      & 1/3 & 80.8 & yes \\
\tableline
\end{tabular}
\tablenotetext{a}{All the microlenses are assumed to have a similar mass of $M = 1 M_\odot$}
\end{center}
\end{table}

\begin{figure*}
\epsscale{0.70} 
\plotone{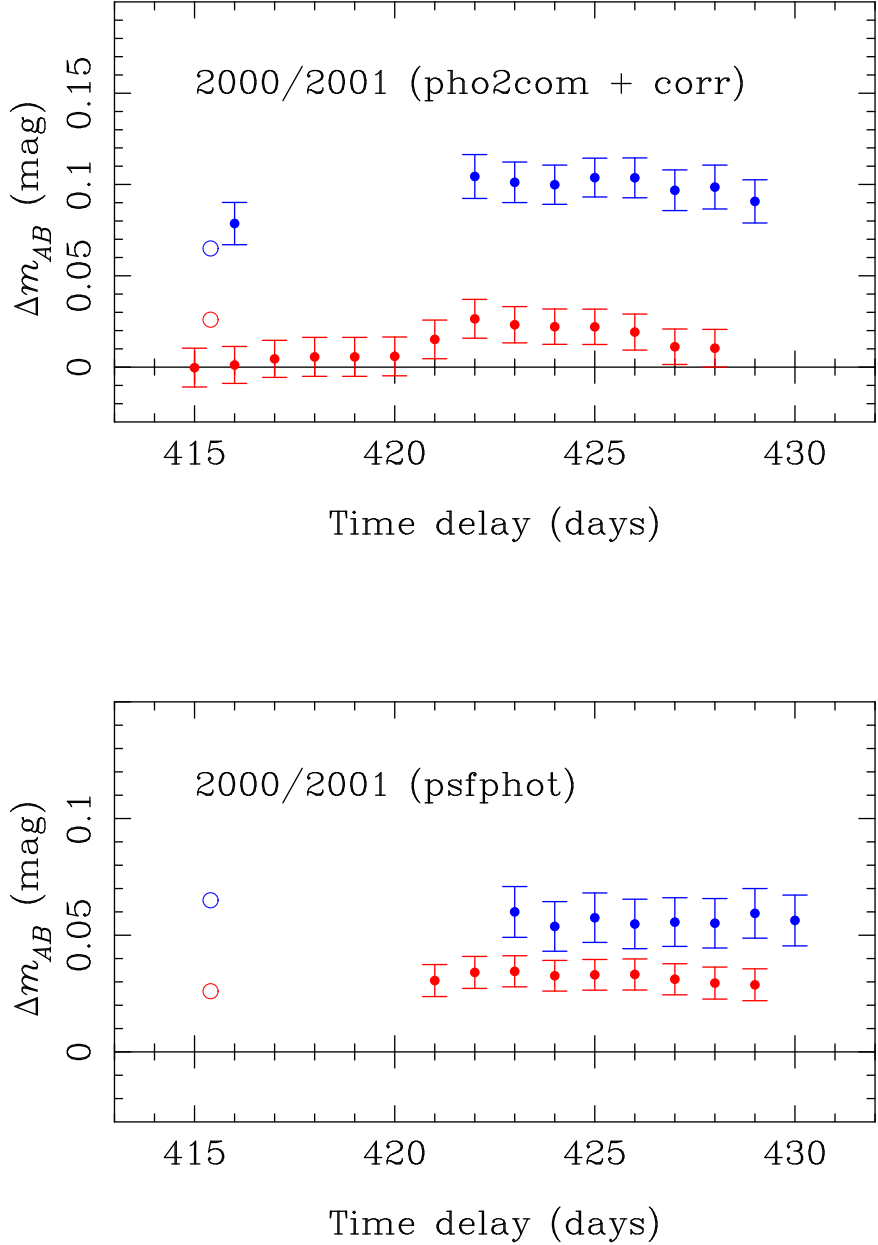}
\caption{QSO 0957+561 magnification ratios in 2000 (2000/2001 seasons). While the top panel 
contains the results from the photometric method {\it pho2comC}, the bottom panel includes the 
results from an alternative photometric task ({\it psfphot}). Blue and red points are 
associated with ratios in the $V$ and $R$ bands, respectively. The filled circles are our 
measurements and the open circles are the typical estimates by the Oslo group (they did not use
{\it pho2comC} or {\it psfphot}, but a different task). We are able to find ratios for time 
delays different to 415--416 days. In particular, the 420--430 days interval is very well
tested.}
\end{figure*}

\begin{figure*}
\epsscale{0.70} 
\plotone{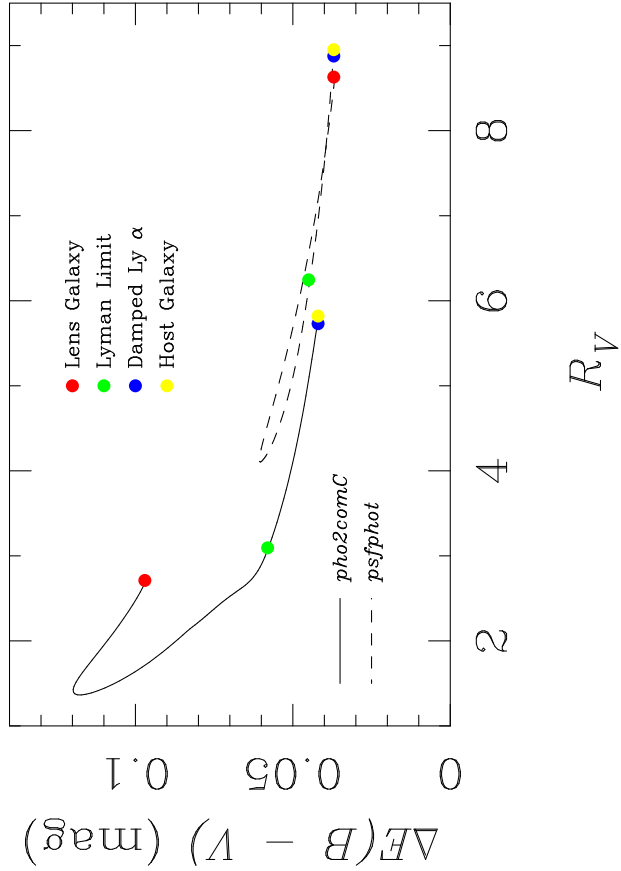}
\caption{$\Delta E(B - V)$ and $R_V$ extinction parameters for dusty objects at 0.36 $\leq 
z_{dust} \leq$ 1.41. The solid line represents the parameters from the typical delay-averaged
{\it pho2comC} ratios, whereas the dashed line traces the parameters from the typical 
delay-averaged {\it psfphot} ratios. The red points are the solutions for a dust system within 
the lens galaxy ($z_{dust}$ = 0.36), the green points represent the solutions for $z_{dust}$ = 
1.1249 (Lyman limit system), the blue points are related to a dusty damped object
($z_{dust}$ = 1.3911), and the yellow points are associated with a dusty host galaxy ($z_{dust}$
= 1.41).}
\end{figure*}

\begin{figure*}
\epsscale{0.70} 
\plotone{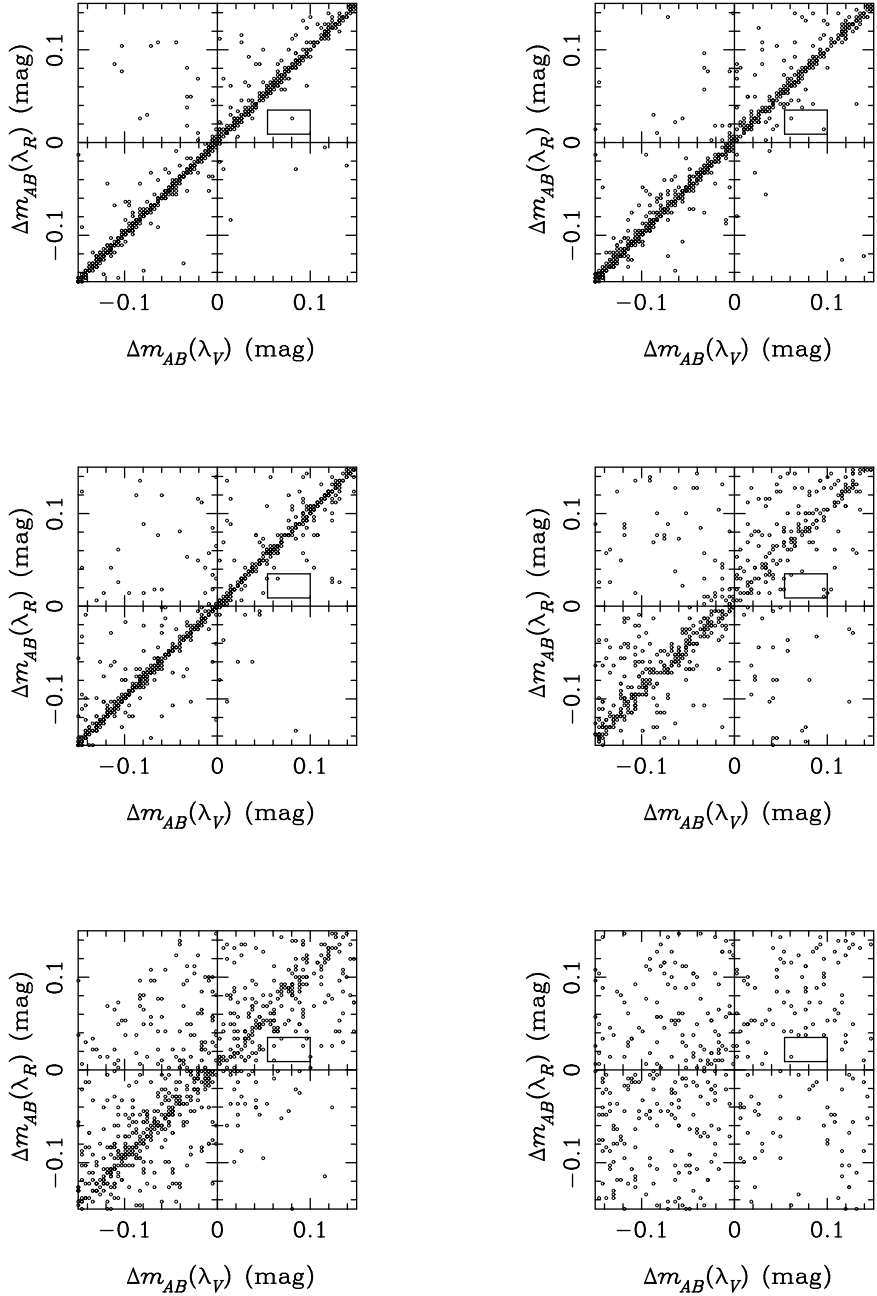}
\caption{Synthetic distributions of [$\Delta m_{AB}(\lambda_V)$,$\Delta m_{AB}(\lambda_R)$] 
pairs for a deflector with only granular matter ($M = 1 M_\odot$) and Gaussian sources. The
radii of the $V$--band sources are $R_V$ = 3 $\times$ 10$^{14}$ cm (top panels), $R_V$ = 10$^{15}$ 
cm (middle panels), and $R_V$ = 3 $\times$ 10$^{15}$ cm (bottom panels). In the left panels, we
consider a standard $q = R_V/R_R$ of about 0.8, and in the right panels, we take a non--standard 
source size ratio of $q$ = 1/3. The 1$\sigma$ measurements in this paper are also depicted in each 
panel (small rectangle). A few simulated pairs are within the 1$\sigma$ rectangle, so they are 
fully consistent with the observations. However, we generate 10$^4$ pairs for each physical picture,
with the vast majority of points out of the observational box, and even out of the $- 0.15 \leq
\Delta m_{AB}(\lambda_V)$ (in mag) $\leq 0.15$ and $- 0.15 \leq \Delta m_{AB}(\lambda_R)$ (in mag) 
$\leq 0.15$ region.}
\end{figure*}

\begin{figure*}
\epsscale{0.70} 
\plotone{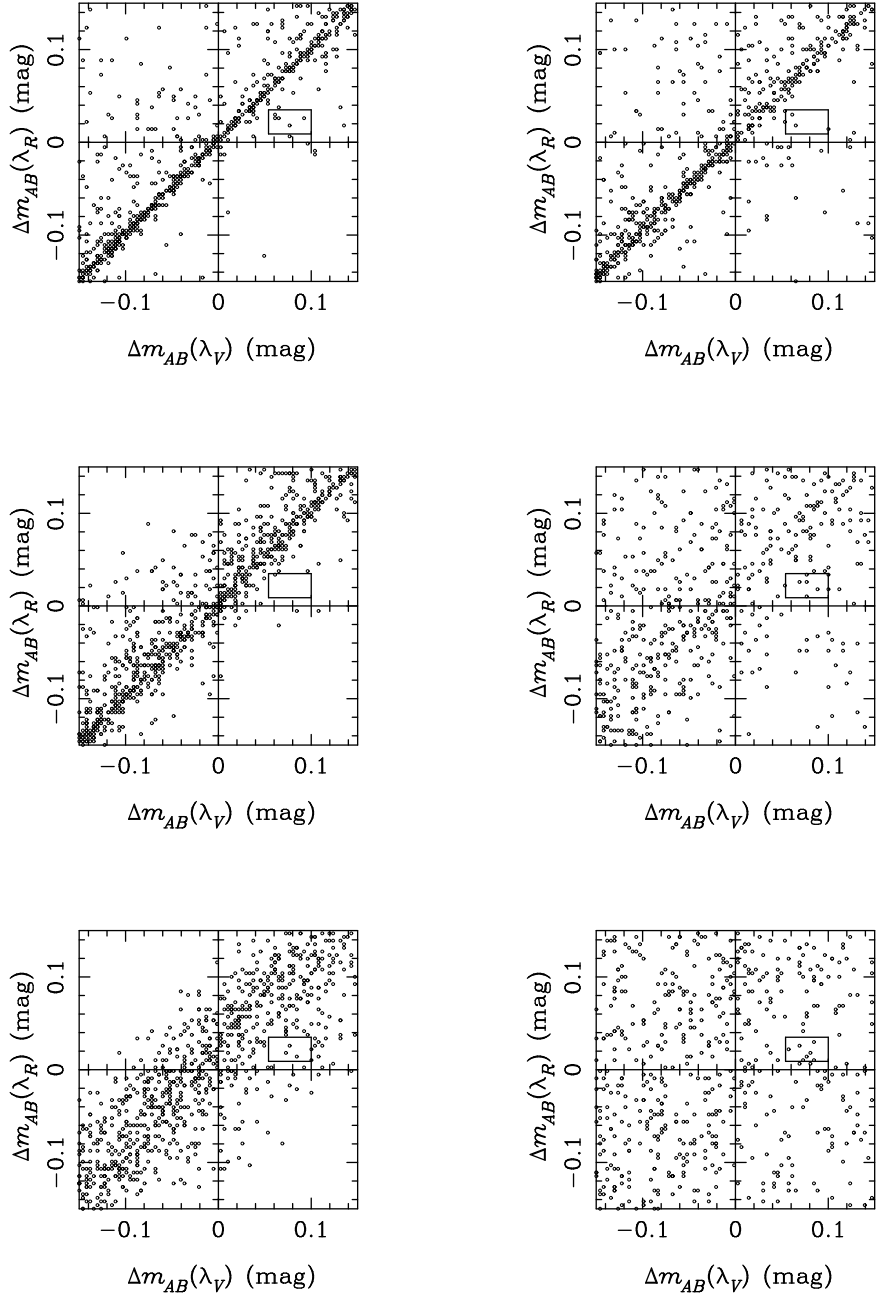}
\caption{Synthetic distributions of [$\Delta m_{AB}(\lambda_V)$,$\Delta m_{AB}(\lambda_R)$] 
pairs for a deflector with only granular matter ($M = 1 M_\odot$) and $p$ = 3/2 power--law 
sources. See the caption in Fig. 3 for comments on the six panels.}
\end{figure*}

\begin{figure*}
\epsscale{0.70} 
\plotone{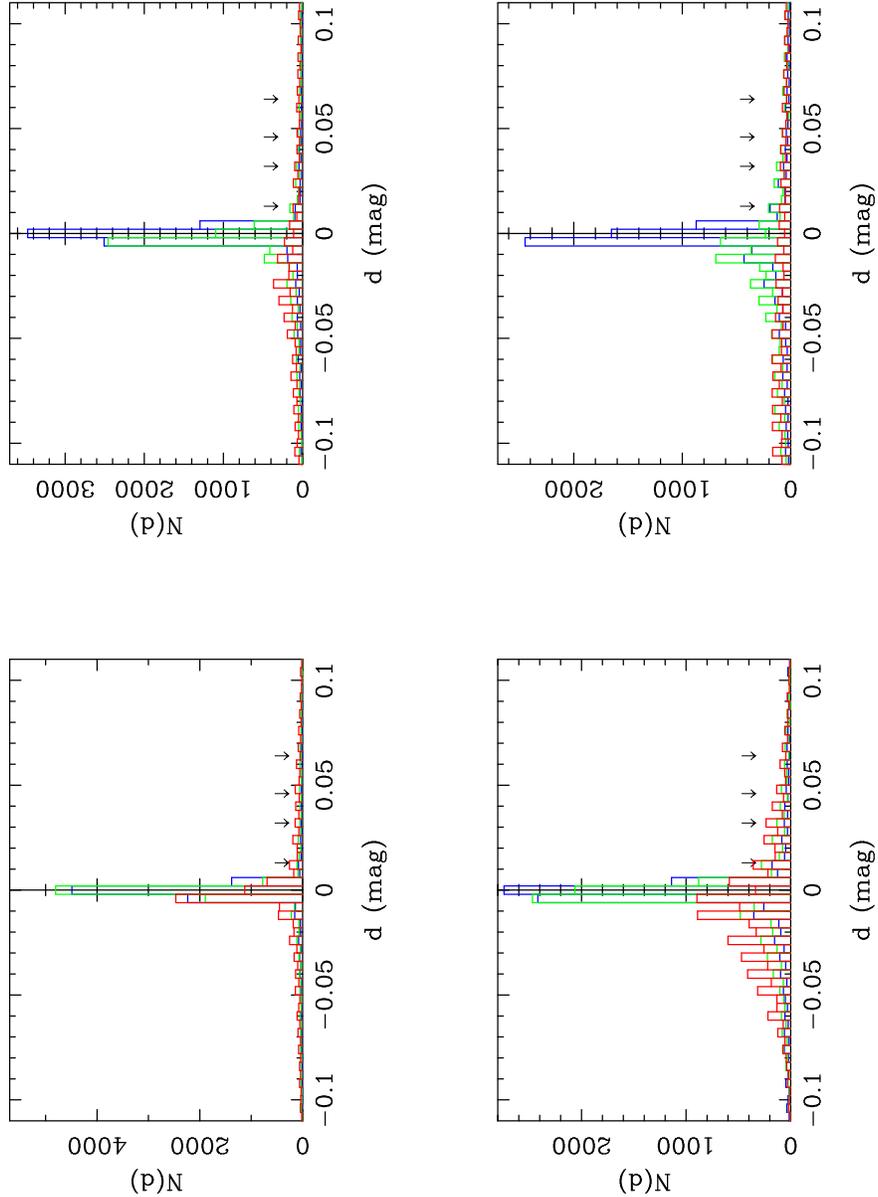}
\caption{Histograms $N(d)$ for the scenarios with 100\% of the mass in compact objects ($M = 1 
M_\odot$). For a given set of physical parameters, we infer a distribution of 10$^4$ points in the 
[$\Delta m_{AB}(\lambda_V)$,$\Delta m_{AB}(\lambda_R)$] plane. In a second step, we obtain the
number of points at different distances $d$ (in magnitudes) from the critical line $\Delta 
m_{AB}(\lambda_R) = \Delta m_{AB}(\lambda_V)$. The distribution $N(d)$ is computed for points above 
the critical line ($d <$ 0) and below the critical line ($d >$ 0). The results from Gaussian sources 
and a standard source size ratio ($\approx$ 
0.8), Gaussian sources and a non--standard $q$ (= 1/3), $p$ = 3/2 power--law sources and a standard 
$q$, and $p$ = 3/2 power--law sources and a non--standard $q$ appear in the top left panel, the top 
right panel, the bottom left panel, and the bottom right panel, respectively. We use blue lines for 
$R_V$ = 3 $\times$ 10$^{14}$ cm, green lines for $R_V$ = 10$^{15}$ cm, and red lines for $R_V$ = 3 
$\times$ 10$^{15}$ cm. The four arrows indicate the distances from the critical line to the four 
vertexes of the 1$\sigma$ rectangle in Figs. 3 and 4.}
\end{figure*}

\begin{figure*} 
\caption{Details on the caustic maps and the locations of the 10$^4$ pairs of pixels used to produce 
Figs. 3--5. The selected pairs of pixels (white marks) are randomly distributed within the 
magnification maps for the A (left panel) and B (right panel) components.}
\end{figure*}

\begin{figure*} 
\caption{Example of chromatic microlensing. We use the magnification maps in Fig. 6 and the largest GS 
sources ($R_V$ = 3 $\times$ 10$^{15}$ cm, $R_R$ = 3 $\times$ $R_V$). The convolved maps for the A
component appear in the top panels: $V$--convolution (left), $R$--convolution (centre), and difference
between both convolutions (right). On the other hand, the convolved maps for the B component are depicted 
in the bottom panels: $V$--convolution (left), $R$--convolution (centre), and difference (right). The 
difference signals have a structure that is correlated to the structure of the caustic networks.}
\end{figure*}


\begin{thebibliography}{}

\bibitem[not]{} Cardelli, J. A., Clayton, G. C., \& Mathis, J. S. 1989, ApJ, 345, 245

\bibitem[not]{} Conner, S. R., Leh\'ar, J., \& Burke, B. F. 1992, ApJ, 387, L61

\bibitem[not]{} Falco, E. E., Impey, C. D., Kochanek, C. S., Leh\'ar, J., McLeod, B. A.,
Rix, H.-W., Keeton, C. R., Mu\~noz, J. A., \& Peng, C. Y. 1999, ApJ, 523, 617

\bibitem[not]{} Garrett, M. A., Calder, R. J., Porcas, R. W., King, L. J., Walsh, D., \&
Wilkinson, P. N. 1994, MNRAS, 270, 457

\bibitem[not]{} Kundi\'c, T., Turner, E. L., Colley, W. N., Gott III, J. R., Rhoads, J. 
E., Wang, Y., Bergeron, L. E., Gloria, K. A., Long, D. C, Malhotra, S., \& Wambsganss, J.
1997, ApJ, 482, 75

\bibitem[not]{} McLeod, B. A., Bernstein, G. M., Rieke, M. J., \& Weedman, D. W. 1998, 
AJ, 115, 1377

\bibitem[not]{} Michalitsianos, A. G., Dolan, J. F., Kazanas, D., Bruhweiler, F. C., 
Boyd, P. T., Hill, R. J., Nelson, M. J., Percival, J. W., \& van Citters, G. W. 1997, 
ApJ, 474, 598

\bibitem[not]{} Motta, V., Mediavilla, E., Mu\~noz, J. A., Falco, E., Kochanek, C. S.,
Garcia--Lorenzo, B., Oscoz, A., \& Serra--Ricart, M. 2002, ApJ, 574, 719

\bibitem[not]{} Mu{\~n}oz, J. A., Falco, E. E., Kochanek, C. S., McLeod, B. A., \& Mediavilla, 
E. 2004, ApJ, 605, 614

\bibitem[not]{} Oscoz, A., Alcalde, D., Serra--Ricart, M., Mediavilla, E., \&  Mu\~noz, 
J. A. 2002, ApJ, 573, L1

\bibitem[not]{} Ovaldsen, J. E., Teuber, J., Schild, R. E., \& Stabell, R. 2003a, A\&A, 
402, 891 

\bibitem[not]{} Ovaldsen, J. E., Teuber, J., Stabell, R., \& Evans, A. K. D. 2003b, MNRAS, 
345, 795

\bibitem[not]{} Pelt, J., Schild, R., Refsdal, S., \& Stabell, R. 1998, A\&A, 336, 829

\bibitem[not]{} Press, W. H., \& Rybicki, G. B. 1998, ApJ, 507, 108

\bibitem[not]{} Refsdal, S., Stabell, R., Pelt, J., \& Schild, R. 2000, A\&A, 360, 10

\bibitem[not]{} Schechter, P. L., Wambsganss, J., \& Lewis, G. F. 2004, astro-ph/0403558

\bibitem[not]{} Schild, R. E., \& Smith, R. C. 1991, AJ, 101, 813

\bibitem[not]{} Schild, R., \& Vakulik, V. 2003, AJ, 126, 689

\bibitem[not]{} Schmidt, R., \& Wambsganss, J. 1998, A\&A, 335, 379  

\bibitem[not]{} Serra--Ricart, M., Oscoz A., Sanch\'\i s, T., Mediavilla, E., Goicoechea, 
L. J., Licandro, J., Alcalde, D., \& Gil--Merino, R. 1999, ApJ, 526, 40 

\bibitem[not]{} Shalyapin, V. N. 2001, Astron. Lett., 27, 150

\bibitem[not]{} Shalyapin, V. N., Goicoechea, L.J., Alcalde, D., Mediavilla, E., Mu\~noz,
J.A., \& Gil--Merino, R. 2002, ApJ, 579, 127

\bibitem[not]{} Smette, A., Surdej, J., Shaver, P. A., Foltz, C. B., Chaffee, F. H., 
Weymann, R. J., Williams, R. E., \& Magain, P. 1992, ApJ, 389, 39

\bibitem[not]{} Stockton, A. 1980, ApJ, 242, L141 

\bibitem[not]{} Ull\'an, A., Goicoechea, L.J., Mu\~noz, J.A., Mediavilla, E., Serra--Ricart, 
M., Puga, E., Alcalde, D., Oscoz, A., \& Barrena, R. 2003, MNRAS, 346, 415

\bibitem[not]{} Walsh, D., Carswell, R. F., \& Weymann, R. J. 1979, Nature, 279, 381

\bibitem[not]{} Wambsganss, J. 1990, PhD thesis, Munich University (report MPA 550)

\bibitem[not]{} Wambsganss, J. 1999, Journ. Comp. Appl. Math., 109, 353

\bibitem[not]{} Wambsganss, J., Schmidt, R. W., Colley, W., Kundi\'c, T., \& Turner, E. 
L. 2000, A\&A, 362, L37

\bibitem[not]{} Wucknitz, O., Wisotzki, L., Lopez, S., \& Gregg, M. D. 2003, A\&A, 405, 445

\bibitem[not]{} Zuo, L., Beaver, E. A., Burbidge, E. M., Cohen, R. D., Junkkarinen, V. 
T., \& Lyons, R. W. 1997, ApJ, 477, 568

\end{thebibliography}
\end{document}